\documentclass[aps,pre,superscriptaddress,floatfix,showpacs,twocolumn,groupedaddress]{revtex4-1}
\usepackage[pdftex]{graphicx}  
\usepackage[update]{epstopdf}
\usepackage{amssymb}
\usepackage{amsmath}
\usepackage{color}
\usepackage{textcomp}
\usepackage[utf8]{inputenc}
\usepackage[normalem]{ulem}
\graphicspath {{./figures/}}
\makeatletter
\def\input@path{{./figures/}}
\makeatother
\usepackage{graphicx}
\usepackage[colorlinks = true,
linkcolor = blue,
urlcolor  = blue,
citecolor = red,
anchorcolor = blue]{hyperref}
\usepackage{amsfonts}
\usepackage{bm}
\usepackage{multirow}
\usepackage{color}
\definecolor{dgreen}{rgb}{0,0.7,0}

\def\bluew#1{{\color{blue} #1}}
\usepackage{tikz}
\usetikzlibrary{shapes,shapes.multipart}

\begin{document}

 \title{Particles confined in arbitrary potentials with a class of finite-range repulsive interactions}
\author{Avanish Kumar}
\affiliation{International Centre for Theoretical Sciences, Tata Institute of Fundamental Research, Bengaluru -- 560089, India}

\author{Manas Kulkarni}\email{manas.kulkarni@icts.res.in}
\affiliation{International Centre for Theoretical Sciences, Tata Institute of Fundamental Research, Bengaluru -- 560089, India}

\author{Anupam Kundu}\email{anupam.kundu@icts.res.in}
\affiliation{International Centre for Theoretical Sciences, Tata Institute of Fundamental Research, Bengaluru -- 560089, India}

\date{\today}

\begin{abstract}

In this paper, we develop a large-$N$ field theory for a system of  $N$ classical particles in one dimension at thermal equilibrium. The particles are 
confined by an arbitrary external potential, $V_\text{ex} (x)$,  and repel each other via a class of pairwise interaction potentials $V_\text{int}(r)$ (where $r$ is distance between a pair of particles) such that 
$ V_\text{int} \sim |r|^{-k}$ when $r \to 0$. We consider the case where every particle is interacting with $d$ (finite range parameter) number of particles to its left and right.  
Due to the intricate interplay between external confinement, pairwise repulsion and entropy, the density exhibits markedly distinct behavior in three regimes $k>0$,~$k \to 0$ and $k<0$.
From this field theory, we compute analytically the average density profile for large $N$ in these regimes. We show that the contribution from interaction dominates the collective behaviour for $k > 0$ and the entropy contribution dominates for $k<0$, and both contributes equivalently in the $k\to 0$ limit (finite range log-gas). 
Given the fact that these family of systems are of broad relevance, our analytical findings are of paramount importance.
These results are in excellent agreement with brute-force Monte-Carlo simulations. 

\end{abstract}


\maketitle

\section{Introduction}

Systems of interacting  particles confined in external potentials is ubiquitous in nature. Particularly, pairwise repulsive interactions with power-law divergences have taken a special place in physics and mathematics. There have been several theoretical investigations on such systems \cite{Forrester_book}. Examples include, one dimensional one-component plasma (1$d$OCP)~\cite{Dhar_2017}, one dimensional coloumb chain~\cite{dubin}, Riesz gas~\cite{Riesz,agarwal_19,Riesz_Hardin2018}, Random Matrix Theory, nuclear physics, mesoscopic transport, quantum chaos, 
number theory~\cite{Mehta_book,Forrester_book,Akemann_book,LMV_book}, Calogero-Moser model~\cite{calogero1969ground,calogero1971solution,calogero1975exactly,moser1976three, polychronakos2006physics, Calogero_1981},  dipolar gas confined to $1d$ \cite{lu2011strongly,Griesmaier_2005,Griesmaier_2006,ni2010dipolar}, screened Coulomb or Yukawa-gas~\cite{Cunden_2017,Cunden_2018} including  finance \cite{Bouchaud_2015} and big-data science \cite{Qiu_2017}. A common feature that most of the above studies have is that the interaction among the particles is long-ranged which means every particle is interacting with every other particle in the system. Such interactions have led to developments of  field theories which have been successfully used to understand various properties like density profiles, number fluctuations, level-spacing distributions, large deviations etc; in equilibrium in the large $N$ limit. In the context of integrable models, such field theories have also been used to understand non-equilirbrium features such as shock waves and solitons~\cite{im1,im2,im3}. 

In most physical systems, however,  interaction between a pair of particles gets often screened which essentially makes the interaction finite-ranged. This naturally raises the following question: What are the effects of finite-ranged interactions on the field theory and the consequences stemming from it?
In this Letter, we precisely address this issue by studying a collection of $N$ classical particles with positions $\{x_i\}$ for $i=1,2,...,N$ in a confining potential $V_\text{ex}(x)$  in one dimension such that $V_{\text{ex}}(x) \to \infty$ as $|x| \to \infty$. Each particle interacts with $d$ particles on its right and left (if available) and they do so via a repulsive interaction $V_\text{int}(r)$ (where $r$ is the distance between a pair of particles) such that $ V_\text{int} \sim |r|^{-k}$ when $r \to 0$ for $k>-k^*$ where $k^*$ is the largest power in the Taylor series expansion of $V_\text{ex}(r)$.  For $k \leq -k^*$, even the ground state (obtained from energy minimisation) of the system is unstable because the particles fly off to  $x=\pm \infty$. It is important to mention that recent cutting-edge developments in experiments has generated a lot of interest in such finite-ranged systems, for e.g., cold atomic gases and ions  \cite{brown2003rotational}, dipolar bosons \cite{lu2011strongly,Griesmaier_2005,Griesmaier_2006, Tang_2018}, Rydberg gases \cite{Jones_2017}.

\section{Model and Properties}
\noindent 
The total energy of our system is given by  
\begin{equation}
E(\{x_i\})= \frac{1}{2} \sum_{i=1}^N V_{\text{ex}}(x_i) + \frac{J~\text{sgn}(k)}{2} \sum_{\substack{|i - j|\leq d \\ j\neq i}} V_{\text{int}} (|x_i-x_j|)
\label{energy_k}
\end{equation}
where $J>0$ and $d$ is an integer. Note that the parameter $d$ in Eq.~\eqref{energy_k} determines the number of particles that each particle is allowed to interact with. For example, by increasing the value of $d$ from $1$ to $N-1$, one can go from nearest neighbour interaction to all-to-all interaction scenario. 
This model is a generalisation of the so called Riesz gas~\cite{Riesz}. Since we are interested in the equilibrium statistical properties of only the position degrees of freedom, the kinetic energy term in the Hamiltonian is omitted.

For the energy in Eq.~\eqref{energy_k}, the equilibrium joint probability distribution function (PDF) of the positions of the particles at finite temperature 
$T=1/\beta$ is given by  $P(x_1,\cdots, x_N) = \frac{1}{Z_N(\beta)} e^{-\beta E[\{ x_i\}]}$, where the partition function $Z_N(\beta)=\int \prod_{i=1}^N {dx_i}\, e^{-\beta E[\{x_i\}]}$. While the confining potential tries to pull all the particles to it's minimum, the pairwise repulsion as well as the entropy tries to spread them apart. Because of this intricate competition, it turns out that the particles settle down over a finite region  $[-\ell_N, \ell_N]$ for $k>0$ and over the whole line for $k\leq0$ with an average macroscopic density 
$\langle \hat{\rho}_N(x)\rangle = N^{-1}\, \sum_{i=1}^N \langle \delta(x-x_i)\rangle$, where $\langle \ldots \rangle$ 
denotes an average with respect to the Boltzmann weight. An important question to ask is: what is the average density for large $N$ and how does it depend on $T,~k$ and $d$? 

\section{Key Findings}

In this paper,  we address the question of average density for $d \sim O(1)$ and find three distinct fascinating scenarios. We show that, for $k>0$ the average density is obtained from a field theory where the interaction term dominates. On the other hand for $k<0$ the entropy dominates. Remarkably, for $k\to 0$, both interaction and entropy contributes equivalently at finite temperature.

In particular, for an external potential of the  polynomial form of $n^{\text{th}}$ order, $V_\text{ex}(x_i) = \sum_{p=1}^{n} a_p x_i^p$, we find that for $k>0$, the average density has the following scaling form $\langle \hat\rho_N(x) \rangle= \ell_N^{-1}~f_k(x/\ell_N)$ in the large $N$ limit where $\ell_N = N^{\frac{k}{k+n}}$ and 
\begin{align}
\begin{split}
f_k(y)=A_d(k)&\left[ 2 \mu_d(k) -V_{\text{ex}}(y)\right]^{1/k},\\
\text{for,} & ~~~|y|\leq \Sigma(\mu_d(k)).
\end{split}
\label{den_k>0}
\end{align}
The edge of the density $\Sigma(\mu_d(k))$ can be obtained from the the real zero, closest to the origin, of the equation $V_{\text{ex}}(y)=2 \mu_d(k)$  and 
$A_d(k)=[2 J \zeta_d(k)(k+1)]^{-\frac{1}{k}}$
with $\zeta_d(k)=\sum_{n=1}^dn^{-k}$.
The function $\mu_d(k)$  can then be determined by the normalisation condition, 
\begin{equation}
\int_{-\Sigma(\mu_d(k))}^{\Sigma(\mu_d(k))} f_k(y) dy=1 \, .
\label{norm_k2_SM2}
\end{equation}
In order to make sure that all terms in the polynomial contribute at an equal footing, the coefficients themselves need to be scaled as $a_{p} \sim N^{\frac{k}{k+n}(n-p)}$. It is important to mention that external potentials in the form of polynomials are of relevance both experimentally as well as theoretically \cite{Polychronakos_1992b,Garraway_2016} and for such potentials we have $k^*=n$. Note that, no such finite bound on $k$ exists for those $V_{\text{ex}}(x)$ which have infinite series representations, for {\it e.g.} box-like potentials such as $V_{\text{ex}}(x)=a~\text{cosh}(b x)$ \cite{Polychronakos_1992a, Gaunt_2013, im3, box1,box2, box3}.


In the $k<0$ case, for any arbitrary external potential $V_{\text{ex}}(x)$, we find that the entropy term dominates to yield\begin{equation}
f_k(y)={e^{-\beta V_{\text{ex}}(y)}}/{C}, \text{ for}~-\infty \leq y \leq \infty, 
\label{kneg}
\end{equation}
with $\ell_N =1$ (no scaling). The normalisation constant $C$ is fixed by $\int_{-\infty}^{\infty} f_k(y) dy=1$.

The $k\to 0$ limit turns out to be very interesting and subtle. To make sense of this limit, we choose $V_{\text{int}}(r)=|r|^{-k}$. Replacing ${\rm sgn}(k)$ in Eq.~(\ref{energy_k}) by $\pm1$ for $k \to 0^{\pm}$, we use $|r|^{-k} \approx 1 - k 
\log |r|$ and set $J=1/|k|$. This up to an overall additive constant provides
\begin{eqnarray}\label{log_gas}
E(\{ x_i\}) = \frac{1}{2} \sum_{i=1}^N V_{\text{ex}}(x_i) - \frac{1}{2} \sum_{\substack{|i - j|\leq d \\ j\neq i}} \ln |x_i-x_j| \;.
\end{eqnarray}
We call this system as the finite-range log-gas \cite{Pandey_2017, Pandey_2019a, Pandey_2019b}. The $k \to 0$ limit can also be taken for some other choices of $V_{\text{int}}(r)$ such as $1/|\sin(r)|^k,~1/|\text{sinh}(r)|^k$, which yields generalised versions of the finite-range log-gas 
where the interaction term inside the summation becomes $\ln |\sin(x_i-x_j)| $ and $ \ln |\text{sinh}(x_i-x_j)|$ respectively \cite{Forrester_book,Mehta_book}.
For all these cases, it turns out that, contributions from both interaction and entropy appear at the same order of $N$ and one finally gets 
\begin{equation}
f_k(y) = e^{-\frac{\beta V_{\text{ex}}(y)}{\beta d+1}} / C_0,  \text{ for } ~-\infty \leq y \leq \infty 
\label{flg}
\end{equation}
with $\ell_N=1$ and $C_0$ being the normalisation constant. 

\begin{figure}[t]
 \includegraphics[width=3.5in]{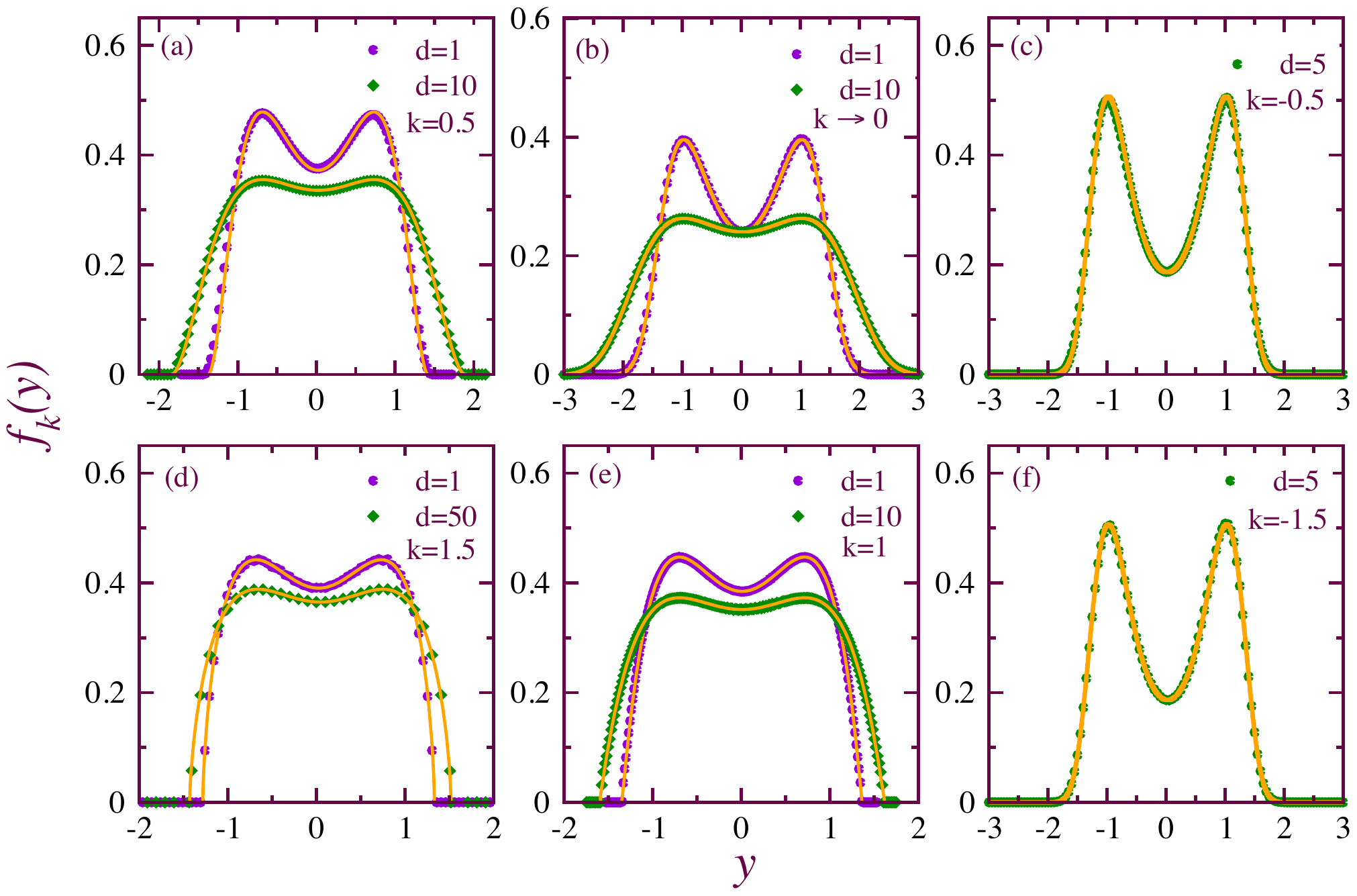}
 \caption{Comparison of the densities with Monte-Carlo simulation for different values of $k$ and $d$. The external potential for all the plots, is
 $V_{\text{ex}}(x)=\frac{1}{2}\left(x^4-N^{\frac{2k}{k+2}}~x^2 \right)$. The interaction potential used in plots (a, c, d, e, f) is $V_{\text{int}}(r)=|r|^{-k}$ whereas in plot (b) it is $V_{\text{int}}(r)=-\ln |r|$. The solid lines in each plot are from theory and symbols are from numerical simulation. For plots with $k>0$ (a, d, e), the theoretical densities are 
given in Eq.~\eqref{den_k>0}. For the Log-gas case (b), $k \to 0$, we compare simulation data with analytical expression in Eq.~\eqref{flg}. The plots (c) and (f) on the right 
column corresponds to $k<0$ where we find Boltzmann-distribution given in Eq.~\eqref{kneg}. Excellent agreement is seen in all cases with no fitting parameters.}
\label{fig1}
\end{figure}

\begin{figure}[t]
 \includegraphics[width=3.5in]{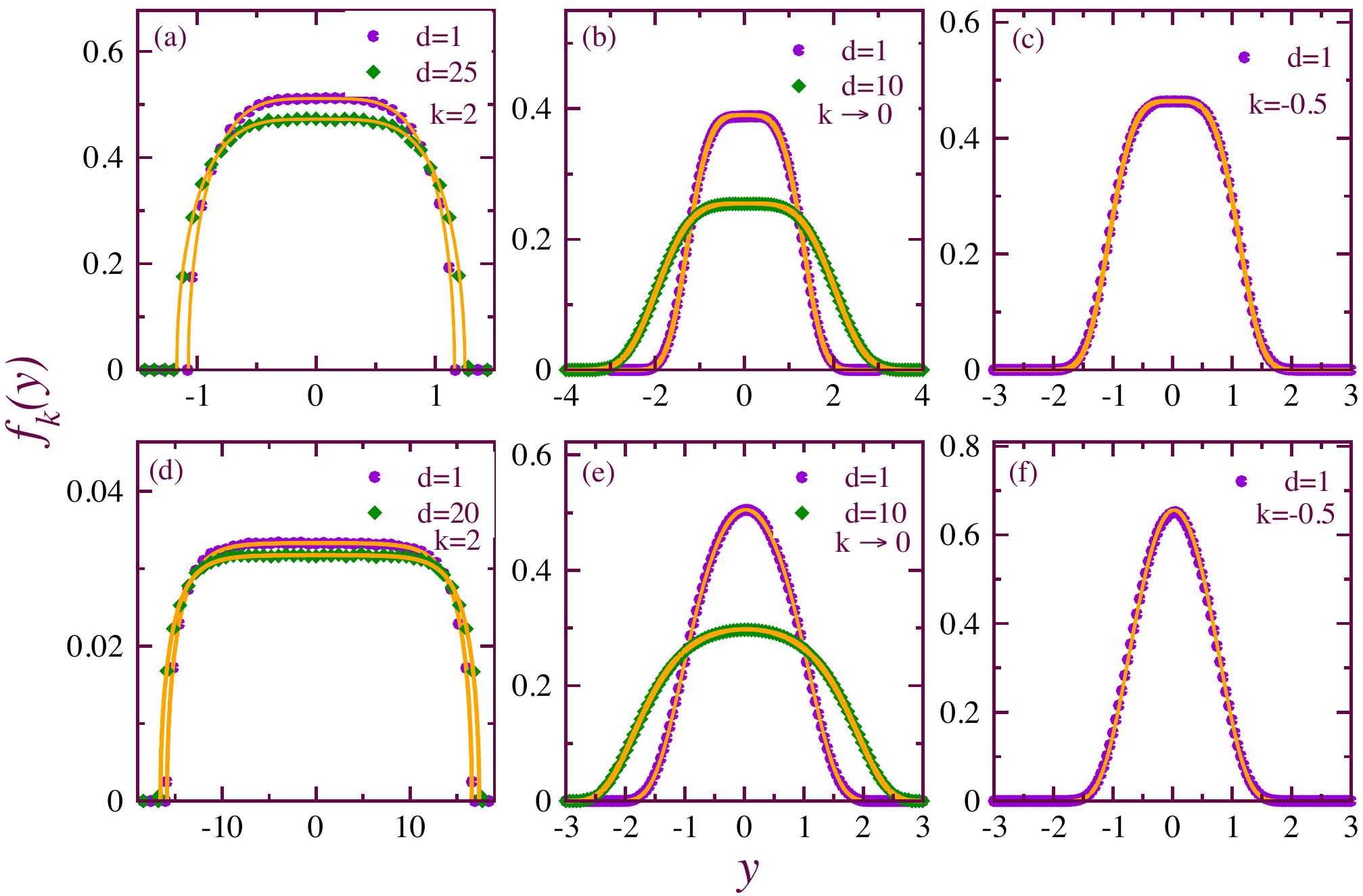}
 \caption{Demonstration of the validity of our theoretical results in more general cases of interactions as well as external potentials. For the plots in the top row, the external potential is $V_{\text{ex}}(x)=x^4/2$. For the plots in the bottom row we have $V_{\text{ex}}(x)=\frac{1}{2}\text{Cosh}\left(\frac{x}{2}\right)$, which naturally sets a $N$ independent length scale {\it i.e.} $\ell_N \sim O(1)$ because it is not in the form a finite-degree polynomial (diverges exponentially at $|x| \to \infty$). The interaction potential used for the  plots in the first column, (a, d) is $V_{\text{int}}(r)=\frac{1}{|\text{Sinh}(r)|^k}$ whereas in the second column,  (b, e) it is $V_{\text{int}}(r)=\ln|\text{Sinh}(r)|$ and in the third column, (c, f) it is $V_{\text{int}}(r)=\sqrt{r}+\frac{r^{5/2}}{12}$ with $r=|x_i-x_j|$. The solid lines in each plot correspond to our theoretical results and the symbols are from numerical simulations. For plot (a), we find that the density is given by Eq.~\eqref{den_k>0}. In plot (d) we compare our simulation data with the analytical expression $f_k(x)= C^{-1/k}(2\mu-N \text{Cosh}(x/2))^{1/k}$ with $C=2 J (k+1)\zeta_d(k)N^{k+1}$, where $\mu$ is fixed by normalisation. The solid lines of the plots in the second and third columns are given in Eqs.~\eqref{flg} and \eqref{kneg}, respectively. Once again we observe excellent agreement with no fitting parameters.}
\label{fig2}
\end{figure}

We also performed Monte-Carlo (MC) simulations for several values of $k$ and find excellent agreement with our analytical predictions (see Fig.~\ref{fig1} and \ref{fig2}). In what follows we discuss the derivation of the large $N$ field theory and the saddle point calculations that lead to our results.

\section{Large-$N$ field theory}

We are interested to compute $\langle \hat{\rho}_N(x)\rangle$ for large $N$ which is formally given by the following functional integral
\begin{align}
\langle \hat{\rho}_N(x)\rangle= \int \mathcal{D}[\rho(z)]~\mathbb{P}[\hat{\rho}_N(z)=\rho(z)]~\rho(x),
\label{av-rho}
\end{align}
$\forall x$, where $\mathbb{P}$ represents the joint probability density functional (JPDF) that $\hat{\rho}_N(z)=\rho(z),~\forall~z \in [-\infty,\infty]$.  
The JPDF, for large $N$, can be written as 
\begin{align}
\mathbb{P}[\hat{\rho}_N(z)&=\rho(z)] =\frac{\int dx_1...dx_N \delta[\hat{\rho}_N(z)-\rho(z)]e^{-\beta E(\{x_i\})}}
{\int dx_1...dx_N e^{-\beta E(\{x_i\})}} \nonumber 
\\&=\frac{ \mathcal{J}_N[\rho(z)] e^{-\beta \mathcal{E}_N[\rho(z)]}\delta \left(\int \rho(z)dz -1\right)}
{\int \mathcal{D}[h(z)] \mathcal{J}_N[h(z)] e^{-\beta \mathcal{E}_N[h(z)]}\delta \left(\int h(z)dz -1\right)} \nonumber 
\end{align}
 where we have assumed that for large $N$, the energy in Eq.~\eqref{energy_k} can be expressed as a functional of the macroscopic density $\hat{\rho}_N(z)=N^{-1}\, \sum_{i=1}^N \delta(z-x_i)$ {\it i.e.} $E(\{x_i\}) \approx \mathcal{E}_N[\hat{\rho}_N(z)]$. In fact this is shown explicitly later [after Eq.~\eqref{av-rho-2}].
The combinatorial factor $\mathcal{J}_N[\rho(z)]$ counts the number of  microscopic configurations compatible with given macroscopic profile $\rho(z)$. In fact  $\mathcal{J}_N[\rho(z)]$ is actually the exponential of the entropy associated to macroscopic density profile $\rho(z)$
\cite{Dean_2008, Dean_2006}
\begin{align}
\mathcal{J}_N[\rho(z)]=e^{\bluew{-}N\int dz~\rho(z)\ln \rho(z)}. \label{entropy}
\end{align}
The delta function $\delta \left(\int \rho(z)dz -1\right)$ ensures the normalisation of the density functions. Replacing this normalisation constraint by its integral representation $\int \frac{d\mu}{2 \pi} e^{-\mu w}=\delta(w)$ (where the integral is along the imaginary $\mu$ axis)  we get 
 \begin{align}
 \mathbb{P}[\rho(z)]&=\frac{\int d\mu~e^{-S_{N,\mu}[\rho(z)]}}{\int d \mu \int \mathcal{D}[h(z)]e^{-S_{N,\mu}[h(z)]}},~~\text{with,} \label{jpdf-2} \\
 \begin{split}
  S_{N,\mu}[\rho(z)]&=\beta \mathcal{E}_N[\rho(z)] \bluew{+} N\int dz \rho(z) \ln \rho(z)  \\ 
 &~~~~~~~~~~~~~~~~~~
 +\mu \left(\int \rho(z)dz -1\right).
  \end{split}
 \label{action-S}
 \end{align}
 We find (shown later) that the functional $S_{N,\mu}[\rho(z)]$, for large $N$ grows as $N^{\gamma_k}$ with $\gamma_k >1$. Hence the partition function in the denominator of the Eq.~\eqref{jpdf-2} can be performed using saddle point 
 method to give 
 \begin{align}
 \mathbb{P}[\rho(z)] \simeq \int d\mu~e^{-\left(S_{N,\mu}[\rho(z)]-S_{N,\mu^*}[\rho^*_N(z)]\right)} \label{jpdf-3}
 \end{align}
 where $\rho^*_N(z)$ and $\mu^*$ are obtained by minimising the action in Eq.~\eqref{action-S} with respect to $\rho(z)$ and $\mu$ {\it i.e} solving the following equations
 \begin{align}
 \frac{\delta S_{N,\mu}[\rho(z)]}{\delta \rho(z)}\Big{|}_{\rho=\rho^*_N}=0,~\text{with}~\int dz~\rho^*_N(z)=1.
 \end{align}
 Using the JPDF $\mathbb{P}$ from Eq.~\eqref{jpdf-3} in Eq.~\eqref{av-rho} and again performing a saddle point integration for large $N$ we find that  the average density profile is same as the most probable or the typical density profile {\it i.e.} 
 \begin{align}
 \langle \hat{\rho}_N(x) \rangle = \rho_N^*(x).\label{av-rho-2}
 \end{align}
 
Next we compute the functional $\mathcal{E}_N[\rho(z)]$ for the energy function given in Eq.~\eqref{energy_k}. To do so we adapt the main idea of Ref.~\onlinecite{agarwal_19}. We first define a smooth function $x(s)$ such that $x(i)=x_i$. This function $x(s)$ becomes unique in the thermodynamic limit \cite{im2} and for a given density profile $\rho(x)$, the position function $x(i)$ is given explicitly by 
\begin{align}
i=N\int^{x(i)}_{-\infty} dz~\rho(z). \label{def-x(i)}
\end{align}
Taking single derivative with respect to $x$ on both sides, we get $di/dx = N\rho(x)$, using which it is easy to see that for any smooth function $g(x_i)$ of the coordinate $x_i$
\begin{align}
\sum_ig(x_i) = N \int dx~g(x) \rho(x). \label{sum-int}
\end{align}
This can be directly applied to the external potential term in Eq.~\eqref{energy_k} to get $\mathcal{E}_N^{\text{ex}}[\rho(x)]= (N/2)\int dx~V_{\text{ex}}(x)\rho(x)$. Expressing the interaction term in terms of the density profile $\rho(x)$ is far from obvious and is discussed below. Using  Eq.~\eqref{def-x(i)}, we write the interaction term in Eq.~\eqref{energy_k} as $\mathcal{E}_{\text{int}}=\sum \sum V_{\text{int}}(|i-j|x'(i)+...)$ where we have used the Taylor series expansion $x(j)=x(i)+(j-i)x'(i)+...$. Assuming, $x'(i)$ is small in the large $N$ limit (see appendices) and using $ V_\text{int}(r)|_{r \to 0} \sim |r|^{-k}$ we get  $\mathcal{E}_{\text{int}}=\frac{J~\zeta_d(k)\text{sgn}(k)}{2} \sum_i (x'(i))^{-k} $ where we have neglected the higher order terms in the Taylor series expansion as they are sub leading.
Now inserting $x'(i)=1/(N\rho(x))$ and using Eq.~\eqref{sum-int} we  have 
\begin{align}
\mathcal{E}_{\text{int}}
&= J~\zeta_d(k)\text{sgn}(k) N^{k+1}\int dx \rho^{k+1}(x).
\end{align}
Hence the total energy functional $\mathcal{E}_N=\mathcal{E}_{\text{ex}}+\mathcal{E}_{\text{int}}$ is given by 
\begin{align}
\begin{split}
\mathcal{E}_N[\rho(x)]&=\frac{N}{2}\int dx~V_{\text{ex}}(x)\rho(x) \\
&~~~~+ J~\zeta_d(k)\text{sgn}(k)~N^{k+1} \int dx \rho^{k+1}(x).
\end{split}
\label{mcal_E}
\end{align}
Following the same procedure it is possible to show from Eq.~\eqref{log_gas} that in the $k \to 0$ one gets
\begin{align}
\begin{split}
\mathcal{E}_N[\rho(x)]&=\frac{N}{2}\int dx~V_{\text{ex}}(x)\rho(x) + Nd \int dx \rho(x) \ln \rho(x).
\end{split}
\label{mcal_E-LG}
\end{align}
This result can also be obtained directly from Eq.~\eqref{mcal_E} in the $k \to 0$ limit after setting $J=1/|k|$.
We now discuss the three regimes separately.

\section{Discussions}

\textit{Regime: $k>0$}: Inserting the above expression of the energy functional in Eq.~\eqref{action-S}, we observe that in the leading order one can neglect the entropy contribution \cite{Lahiri_2000}. For an external potential of $n$th order polynomial form, minimizing this action one finds that $\langle \hat\rho_N(x) \rangle= \rho_N^*(x) = \ell_N^{-1}~f_k(x/\ell_N)$ with $\ell_N =N^{\frac{k}{k+n}}$ and $f_k(y)$ given in Eq.~\eqref{den_k>0}. This result is verified numerically. Using this scaling form of the density in the action in Eq.~\eqref{action-S} back, it is easy to see that $\mathcal{S}_{N,\mu^*} \sim N^{\gamma_k}$ with $\gamma_k=\frac{k(n+1)+n}{k+n}$. 
In fact for $k>1$, the formula in Eq.~\eqref{den_k>0}, holds for any $d$ even when $d=N-1$ for which $\zeta_d(k)$ becomes the usual Riemaan zeta function $\zeta(k)$. This happens because, for $k>1$, the contribution from all-to-all interaction comes only at $ O (N^2)$ which is still subdominant \cite{agarwal_19}. It is to be noted that the above analysis fails for very high temperatures of the order $\sim O(N^{-\frac{2k}{k+n}})$ when entropy becomes important. 
In the special case of a quadratic potential (i.e., $n=2$ with, $a_1=0$ and $a_2=1$), we get the following result for the support, 
$\mu_d(k)=\frac{1}{8}[A_d(k) B(1+1/k,1+1/k)]^{-\frac{2k}{k+2}}$
and $\Sigma(\mu_d(k))= \sqrt{2\mu_d(k)} $.

\textit{Regime 2: $k < 0$}: In this regime, interestingly, as pairwise interaction is of $O(N^{1-|k|})$, it becomes irrelevant 
in comparison to the entropy term which is of $O(N)$. Therefore,  minimizing the action which now involves only the external potential and entropy gives us the usual Boltzmann distribution in Eq.~\eqref{kneg} with $\ell_N=1$ for any external potential. It is noteworthy, that the density profile becomes independent of the details of the interaction although it plays important role to have a description in terms of macroscopic particle densities. 

\textit{Regime 3: $k\to 0$}: In the case of finite range log-gas, as can be seen using Eq.~\eqref{mcal_E-LG} in Eq.~\eqref{action-S}, there is an intricate interplay between pairwise interaction and entropy because they contribute at the same order.  Minimizing this action Eq.~\eqref{action-S}, we get Eq.~\eqref{flg} for any external potential. This result was also recently obtained via a microscopic method~\cite{Pandey_2019a}.  Interaction energy and entropy contributing equivalently has also been observed in log gas with all-to-all interactions \cite{Allez_2012,Allez_2013}.

\section{Numerical method and details}   

Our analytical predictions were tested against brute-force MC simulations for $N=501$ and $\beta=1$.  
In our simulations we collect data after every 10 MC cycles and averages were performed over around $10^{7}-10^{8}$ samples to compute the particle densities in different cases discussed above. We compare these results with our theoretical expression in Fig.~\ref{fig1} and Fig.~\ref{fig2} and observe excellent agreement in all cases.
To make sure that we collect data after the system has relaxed to equilibrium state, we checked for the
equipartition by computing virial $\langle x_j \frac{\partial E(\{x_i\})}{\partial x_j} \rangle$.  
The excellent agreement with equipartition thereby benchmarking our numerics is given in the appendix.

\section{Conclusions and Outlook} 
In this paper, we derive a large-$N$ field theory for a system of $N$ particles repulsively interacting over a finite-range and  confined in arbitrarily external potentials. We discuss a family of interaction potentials $V_{\text{int}}(r)$ such that they behave as $\sim 1/|r|^k$ for small $r$. We identify three distinct regimes depending on the value of $k$ and for each regimes we derive the action in the large $N$ limit. Minimising this action provides us explicit expressions of the densities in arbitrary confining potentials. 
 Our analytical results of densities are in excellent agreement with our brute-force numerical simulations. 
 It is pertinent to mention that such densities of finite-ranged systems can be experimentally observed in a broad range of experiments such as ions \cite{qubit, srep}, dusty plasma \cite{dp} to name a few. 
 This work is of paramount importance since it is essentially a starting point for any analysis on a broad class of interacting classical systems. 
 For e.g., if one wants to study nonlinear hydrodynamics \cite{Spohn_2014,Kulkarni_2012}, interacting over-damped Langevin particles \cite{Dean_1996}, single-file motion \cite{Demery_2014}, large-deviations \cite{DM_2006, Dean_2008, Dean_2006, Kundu_2016, Dhar_2018} then writing a large-N field theory is the very beginning step and a correct form of the energy functional is crucial.

Our work paves the path for several future studies such as  non-trivial extension to higher dimensions, extreme value statistics, level spacing distributions (i.e. statistics of gap between successive particles) and large deviation functions of these externally confined pairwise interacting particles. Our work acts as a genesis and provides  foundation for embarking on these exciting directions. Furthermore, connections between these models and random matrix theories remains an open and interesting question. Finally, it would also be interesting to understand the crossover from finite-ranged interaction to all-to-all coupling \cite{agarwal_19}. 

\section{Acknowledgments} 

MK would like to acknowledge support from the project 6004-1 of the 
Indo-French Centre for the Promotion of Advanced Research (IFCPAR),  
the Ramanujan Fellowship SB/S2/RJN-114/2016, the SERB Early Career Research Award ECR/2018/002085 and Matrics Grant (MTR/2019/001101) from 
the Science and Engineering Research Board, Department of Science and Technology, Government of India.
AK would like to acknowledge support from the project 5604-2 of the Indo-French Centre for the 
Promotion of Advanced Research (IFCPAR) and the the SERB Early Career Research Award ECR/2017/000634 from  from the Science and Engineering Research Board, Department of Science and Technology, Government of India. We gratefully acknowledge Hemanta Kumar G. and ICTS-TIFR high performance computing facility. 

\appendix


\section{Continuum approximation for the finite-range interaction term}

\begin{figure}
\includegraphics[width=3.75in]{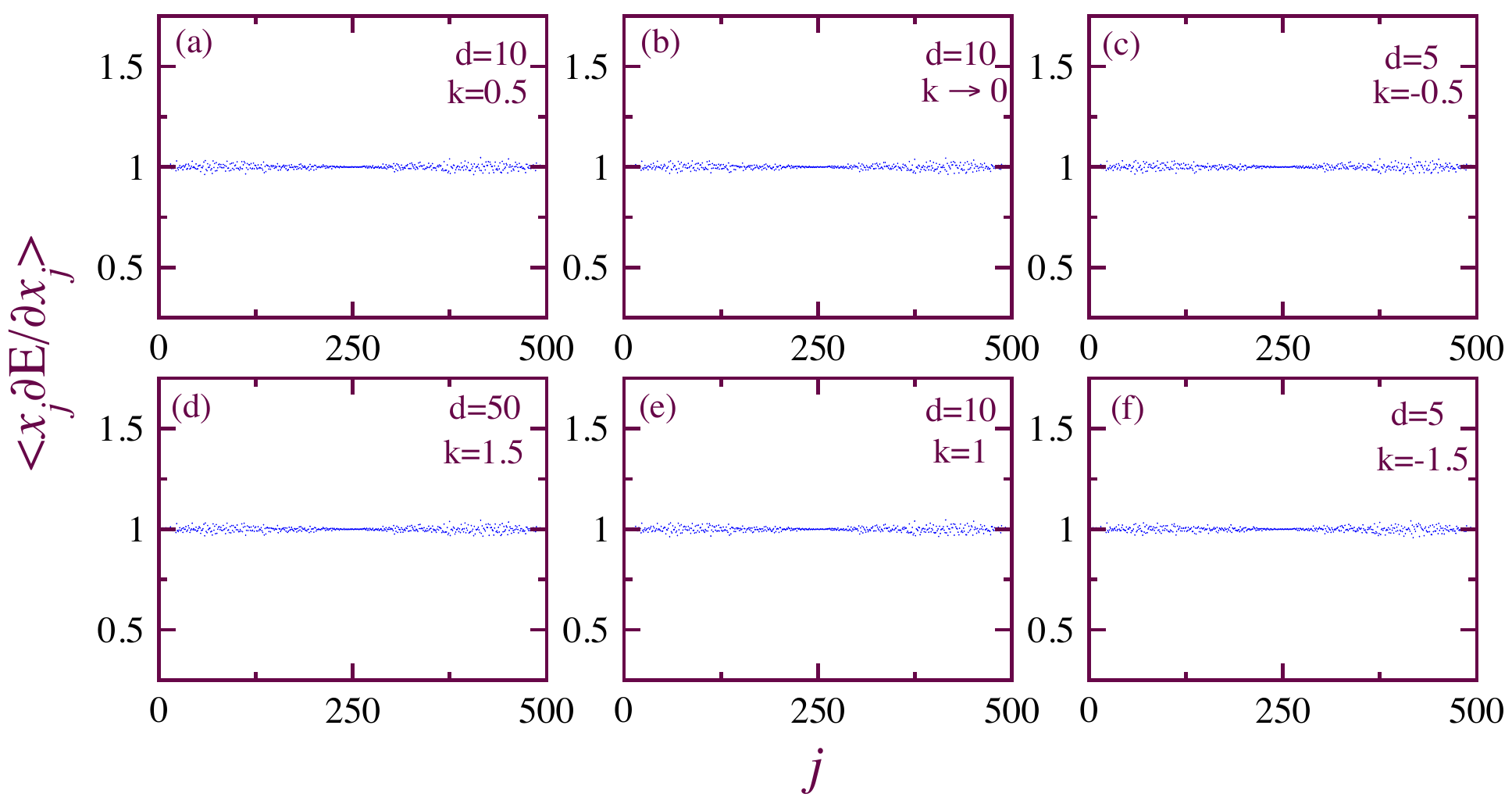}
 \caption{Figures above demonstates the virial plots obtained from Eq.~\eqref{virial-supp} for the corresponsing plots in Fig.~(1) of main text.}
\label{S1}
\end{figure}

\begin{figure}
 \includegraphics[width=3.75in]{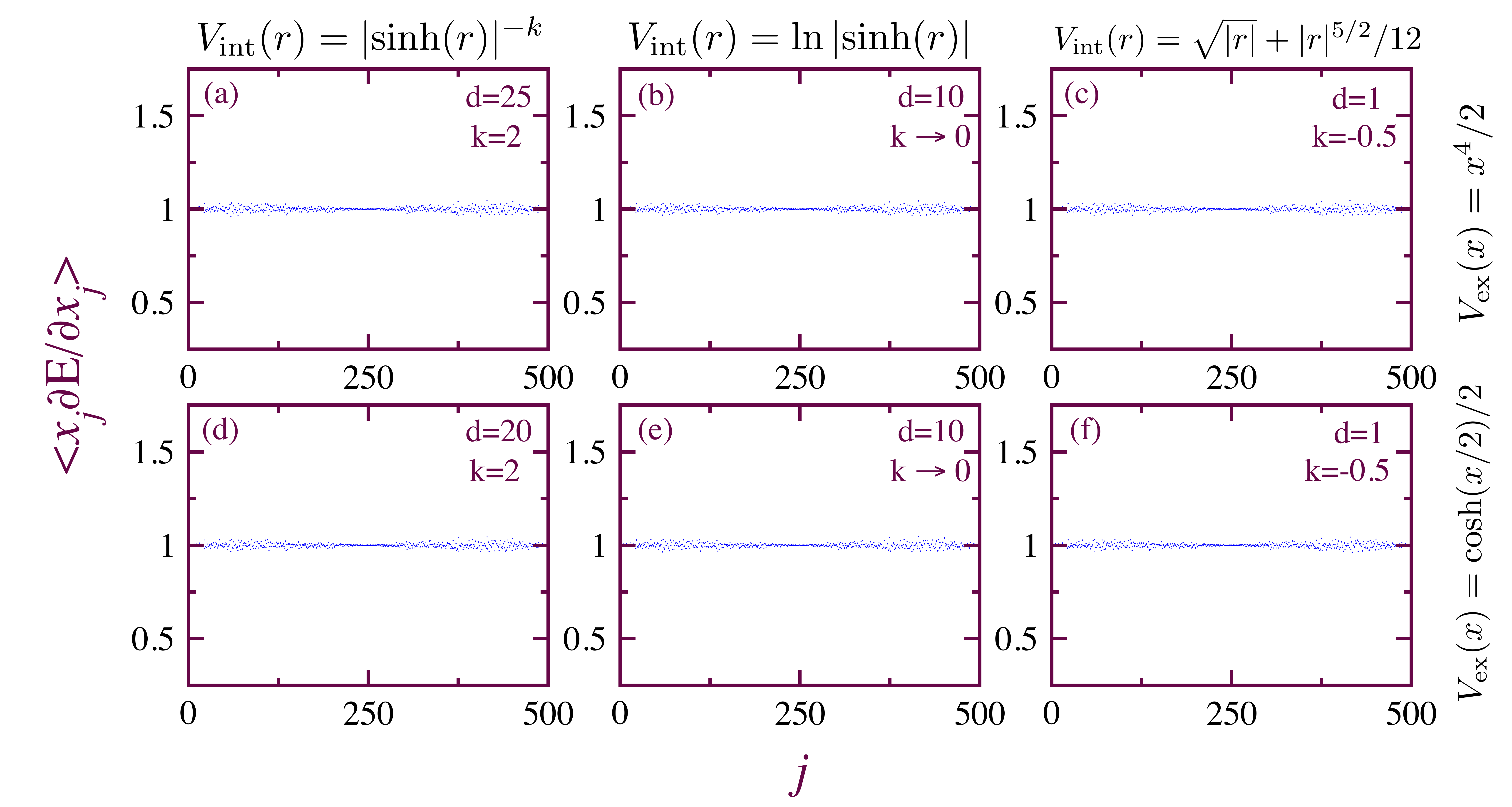}
\caption{Figures above demonstates the virial plots obtained from Eq.~\eqref{virial-supp} for the corresponsing plots in Fig.~(2) of main text.}
\label{S2}
\end{figure}

In Eq. (1) of the main text, we defined the energy of a microscopic configuration $\{x_i\}$ as
\begin{equation}
E(\{x_i\})= \frac{1}{2} \sum_{i=1}^N V_{ex}(x_i) + \frac{J~\text{sgn}(k)}{2} \sum_{\substack{|i - j|\leq d \\ j\neq i}} V_{int} (|x_i-x_j|).
\label{energy_k-sm}
\end{equation}
where $J>0$ and $d$ is an integer. We want to express the interaction term 
\[ \mathcal{E}_{\text{int}}=\frac{J~\text{sgn}(k)}{2} \sum_{\substack{|i - j|\leq d \\ j\neq i}} V_{int} (|x_i-x_j|)\]
as a functional of the macroscopic density $\rho(z)$. As noted in the main text, for large $N$ one can define a smooth function $x(i)$ such that 
\begin{align}
i=N\int^{x(i)}_{-\infty} dz~\rho(z). \label{def-x(i)-sm}
\end{align}
Using this equation, we can write 
\begin{align}
 \mathcal{E}_{\text{int}}=\frac{J~\text{sgn}(k)}{2} \sum_{\substack{|i - j|\leq d \\ j\neq i}} V_{int} \left(\Big{|}\sum_{n=1}^\infty\frac{(i-j)^n}{n!}x^{[n]}(i)\Big{|}\right),
 \end{align}
where $x^{[n]}(i)=\frac{d^n x(i)}{d i^n}$. It easy to see (also justified later) that $\frac{|x^{[n+1]}(i)|}{|x^{[n]}(i)|}\sim O(1/N)$. Hence keeping only the leading order term in the Taylor series expansion in the argument of $V_{\text{int}}$ we have 
\begin{align}
 \mathcal{E}_{\text{int}} &\sim \frac{J~\text{sgn}(k)}{2} \sum_{i=1}^N \sum_{\substack{|i - j|\leq d \\ j\neq i}} V_{int} \left(|i-j|x^{[1]}(i)\right), \nonumber \\
 & \sim \frac{J~\text{sgn}(k)}{2} \sum_{i=1}^N\sum_{\substack{|i - j|\leq d \\ j\neq i}} V_{int} \left(\frac{|i-j|}{N\rho(x(i))}\right), \nonumber \\ 
  & \qquad \qquad \qquad ~~~~~~~~~\text{using}~~x^{[1]}(i)=\frac{1}{N\rho(x(i))} \nonumber \\
 &\sim  \frac{J~\text{sgn}(k)}{2} \sum_{i=1}^N\sum_{\substack{|i - j|\leq d \\ j\neq i}}  \frac{N^k\rho(x(i))^k}{|i-j|^k},
  \label{e_int-1st}
 \end{align}
In the last step we used the fact the $x^{[1]}(i)$ is small, which is because of the following. We expect that $\rho(x)$ should have the following scaling form 
\begin{align}
\rho(x) = \frac{1}{\ell_N}f\left(\frac{x}{\ell_N}\right),~~\text{with},~~\lim \limits_{N \to \infty} \frac{\ell_N}{N} \to 0. \label{rho_lim}
\end{align}
Assuming that  limit in Eq.~\eqref{rho_lim} is true we proceed and compute $\rho(x)$ performing the action minimisation procedure explained in the main text and finally check that this assumption is indeed true ---- thereby making the whole argument self consistent.

\noindent
Simplifying Eq.~\eqref{e_int-1st} further we get 
\begin{align}
 \mathcal{E}_{\text{int}} &\sim  \frac{J~\text{sgn}(k)}{2} \sum_{i=1}^N N^k\rho(x(i))^k \sum_{\substack{|i - j|\leq d \\ j\neq i}}   \frac{1}{|i-j|^k}, \nonumber \\ 
 &\sim  \frac{J~\text{sgn}(k)}{2} \sum_{i=1}^N N^k\rho(x(i))^k \sum_{n=1}^d   \frac{1}{n^k}, \nonumber \\
 &\sim  J~\text{sgn}(k) \zeta_d(k) \sum_{i=1}^N N^k\rho(x(i))^k,  \nonumber \\
 &\sim  J~\text{sgn}(k)~\zeta_d(k)~N^{k+1}\int dx \rho(x)^{k+1}, \label{e_int-2nd} \\ 
 & \qquad\quad\qquad\quad ~~~ \text{using}~~\sum_ig(x_i) = N \int dx~g(x) \rho(x) \nonumber
 \end{align}
The above calculation is true for $k>-k^*$ (see main text). However for $k \to 0$ (finite-range log-gas) the above expression 
gets simplified as follows: Setting $J =1/|k|$ and writing $(N\rho(x))^k = e^{k \ln (N\rho(x))}$ and finally  taking the $k \to 0$, we obtain 
\begin{align}
 \mathcal{E}_{\text{int}} 
 &\sim  d~N\int dx~\rho(x) \ln \rho(x) \label{e_int-lg}
 \end{align}
upto an overall additive constant where we have use $\zeta_d(0)=d$. Now adding this functional form of $\mathcal{E}_{\text{int}}[\rho(z)]$ 
to $\mathcal{E}_{\text{ex}}[\rho(z)]$ we 
get the total energy functional $\mathcal{E}_{N}[\rho(z)]$ for $k \neq 0$

\begin{align}
\begin{split}
\mathcal{E}_N[\rho(x)]&=\frac{N}{2}\int dx~V_{\text{ex}}(x)\rho(x) 
 \\ &+ J~\zeta_d(k)\text{sgn}(k)~N^{k+1} \int dx \rho^{k+1}(x).
\end{split}
\end{align}
Following a similar calculation one can show that the energy functional for $k \to 0$ becomes
 \begin{align}
\begin{split}
\mathcal{E}_N[\rho(x)]&=\frac{N}{2}\int dx~V_{\text{ex}}(x)\rho(x) + Nd \int dx \rho(x) \ln \rho(x).
\end{split}
\end{align}
Inserting these expressions of the energy functionals, in the expression of the action $S_{N,\mu}[\rho(z)]$ below 
\begin{align}
  S_{N,\mu}[\rho(z)]&  =\beta \mathcal{E}_N[\rho(z)] + N\int dz \rho(z) \ln \rho(z)   \nonumber \\ 
  &\qquad \quad +\mu \left(\int \rho(z)dz -1\right).
 \label{action-S-sm}
 \end{align}
and minimizing it we get the following saddle point equations 
 \begin{align}
\frac{1}{2}V_{\rm ex}(x) + J~(k+1) \zeta_d(k)\text{sgn}(k)~N^{k} \rho^k(x) + \mu &=0,\nonumber \\ 
&\text{for}~k>0, \\
\frac{1}{2}V_{\rm ex}(x) + (\beta d+1) [ \ln \rho(x) -1] +\mu=0,\nonumber \\ 
&\text{for}~k=  0,  \\
\frac{1}{2}V_{\rm ex}(x) + [ \ln \rho(x) -1] +\mu=0,\nonumber \\ 
&\text{for}~k<  0, 
 \end{align}
 in the leading order in $N$. Solving these equations we 
find that the saddle point density is given by Eq.~\eqref{rho_lim} with 
\begin{align}
\ell_N = \begin{cases}
&N^{\frac{k}{k+n}},~~\text{for},~k>0 \\
& 1,~~~~~~~~\text{for},~k\le 0
\label{case}
\end{cases}
\end{align}
and 
\begin{align}
f_k(y)=
\begin{cases}
&A_d(k)\left[ 2 \mu_d(k) -V_{\text{ex}}(y)\right]^{1/k},~|y|\leq \Sigma(\mu_d(k)), \\ &\qquad \qquad ~~~~~~~~~~~~~~~~~~~~~~~~~~~~~~~\text{for}~~k>0, \nonumber \\
& \\
& e^{-\frac{\beta V_{\text{ex}}(y)}{\beta d+1}} / C_0,  ~-\infty \leq y \leq \infty ,~~~~\text{for}~~k=0, \\
& \\
&{e^{-\beta V_{\text{ex}}(y)}}/{C}, ~-\infty \leq y \leq \infty, ,~~~~\text{for}~~k<0
\end{cases}
\end{align}
as announced in Eqs.~(2), (6) and (4) in the main text. Here $C$ and $C_0$ are normalisation constants. This clearly justifies the limit in Eq.~\eqref{rho_lim}. It is easy to see that for box-like this limit is trivially true since there is a length scale set by the potential itself.

It is important to note that Eq.~\eqref{case} holds only when the system is stable and there is a notion of a density profile, i.e., Eq.~\eqref{rho_lim}. The most suitable way to visualize this is as follows. If $k<k^*$, then even for a finite number of particles $N$, there is no finite solutions for the particle positions that minimise the energy in Eq.~\eqref{energy_k-sm}. All particles in such a scenario fly away to $\pm \infty$ making the discussion on density to be void.

\section{Virials (equipartition)}

To make sure that we collect data after the system has relaxed to equilibrium state, we checked for the
equipartition by computing virial $\langle x_j \frac{\partial E(\{x_i\})}{\partial x_j} \rangle$. 
Below we show the virials for all the plots in Figs.~(1) and (2) in the main text. The equipartition was tested by checking, 
\begin{equation}
\left\langle x_j \frac{\partial E(\{x_i\})}{\partial x_j} \right\rangle = k_B T
\label{virial-supp}
\end{equation}  
Fig.~\ref{S1} and Fig.~\ref{S2} shows remarkable agreement thereby validating all our numerical results.

Note that for $k>0$ when $T\sim O(1)$, then the finite temperature results match with the density profile obtained by minimizing the total energy Eq.~(1) in the main text. In other words, we have $N$ equations $\frac{\partial E(\{x_i\})}{\partial x_i} = 0,~~i=1,2,...,N$ from which can solve for the $N$ unknows $\{x_i^\text{min};~i=1,2,...,N\}$. Reconstructing a density function from this (say, by using inverse of interparticle distance) will give a density profile which also will agree with the one obtained from minimization of the action described above. This in turn is in perfect agreement with brute force finite temperature Monte-Carlo. Needless to mention, this of-course does not encode any information about fluctuations.


\begin{thebibliography}{100}

\bibitem{Forrester_book} P. J. Forrester, {\it Log-gases and random matrices} (Princeton University Press, Princeton, NJ 2010).

\bibitem{Dhar_2017}A. Dhar, A. Kundu, S.N. Majumdar, S. Sabhapandit, G. Schehr, 
Phys. Rev. Lett.  \textbf{119}, 060601 (2017).

\bibitem{dubin} Daniel H. E. Dubin, 
Phys. Rev. E.  \textbf{55}, 4017 (1996).

\bibitem{Riesz} M. Riesz, Acta Sci. Math. Univ. Szeged, {\bf 9}, 1 (1948).

\bibitem{agarwal_19} S. Agarwal, A. Dhar, M. Kulkarni, A. Kundu, 
S. N. Majumdar, D. Mukamel and G. Schehr, 
Phys. Rev. Lett. \textbf{123}, 100603 (2019)

\bibitem{Riesz_Hardin2018} Hardin, D.P., Leblé, T., Saff, E.B. et al. Constr Approx (2018) 48: 61. https://doi.org/10.1007/s00365-018-9431-9 



\expandafter\ifx\csname natexlab\endcsname\relax\def\natexlab#1{#1}\fi
\expandafter\ifx\csname bibnamefont\endcsname\relax
  \def\bibnamefont#1{#1}\fi
\expandafter\ifx\csname bibfnamefont\endcsname\relax
  \def\bibfnamefont#1{#1}\fi
\expandafter\ifx\csname citenamefont\endcsname\relax
  \def\citenamefont#1{#1}\fi
\expandafter\ifx\csname url\endcsname\relax
  \def\url#1{\texttt{#1}}\fi
\expandafter\ifx\csname urlprefix\endcsname\relax\def\urlprefix{URL }\fi
\providecommand{\bibinfo}[2]{#2}
\providecommand{\eprint}[2][]{\url{#2}}


\bibitem{Mehta_book} M. L. Mehta, {\it Random Matrices} (Academic Press, Amsterdam, 2004).


\bibitem{Akemann_book} {\it The Oxford Handbook of Random Matrix Theory}\,  ed. by G. Akemann, G. Baik, and
P. Di Francesco (Oxford University Press, Oxford, UK, 2011).

\bibitem{LMV_book} G. Livan, M. Novaes, and P. Vivo, {\it Introduction to Random Matrices - Theory and Practice}
(Springer, New York, 2018).


\bibitem[{\citenamefont{Calogero}(1969)}]{calogero1969ground}
\bibinfo{author}{\bibfnamefont{F.}~\bibnamefont{Calogero}},
  \bibinfo{journal}{J. Math. Phy.}
  \textbf{\bibinfo{volume}{10}}, \bibinfo{pages}{2197} (\bibinfo{year}{1969}).


\bibitem[{\citenamefont{Calogero}(1971)}]{calogero1971solution}
\bibinfo{author}{\bibfnamefont{F.}~\bibnamefont{Calogero}},
  \bibinfo{journal}{J. Math. Phys.}
  \textbf{\bibinfo{volume}{12}}, \bibinfo{pages}{419} (\bibinfo{year}{1971}).

\bibitem[{\citenamefont{Calogero}(1975)}]{calogero1975exactly}
\bibinfo{author}{\bibfnamefont{F.}~\bibnamefont{Calogero}},
  \bibinfo{journal}{Lett. Nuovo Cimento}
  \textbf{\bibinfo{volume}{13}}, \bibinfo{pages}{411} (\bibinfo{year}{1975}).

\bibitem[{\citenamefont{Moser}(1976)}]{moser1976three}
\bibinfo{author}{\bibfnamefont{J.}~\bibnamefont{Moser}}, in
  \emph{\bibinfo{booktitle}{Surveys in Applied Mathematics}}
  (\bibinfo{publisher}{Elsevier}, \bibinfo{year}{1976}), pp.
  \bibinfo{pages}{235--258}.

\bibitem{Calogero_1981} F. Calogero, J. Math. Phys. {\bf 22}, 919 (1981).

\bibitem[{\citenamefont{Polychronakos}(2006)}]{polychronakos2006physics}
\bibinfo{author}{\bibfnamefont{A.~P.} \bibnamefont{Polychronakos}},
  \bibinfo{journal}{J. Phys. A: Math. Gen.}
  \textbf{\bibinfo{volume}{39}}, \bibinfo{pages}{12793} (\bibinfo{year}{2006}).


\bibitem[{\citenamefont{Lu et~al.}(2011)\citenamefont{Lu, Burdick, Youn, and
  Lev}}]{lu2011strongly}
\bibinfo{author}{\bibfnamefont{M.}~\bibnamefont{Lu}},
  \bibinfo{author}{\bibfnamefont{N.~Q.} \bibnamefont{Burdick}},
  \bibinfo{author}{\bibfnamefont{S.~H.} \bibnamefont{Youn}}, \bibnamefont{}
  \bibinfo{author}{\bibfnamefont{B.~L.} \bibnamefont{Lev}},
  \bibinfo{journal}{Phys. Rev. Lett.} \textbf{\bibinfo{volume}{107}},
  \bibinfo{pages}{190401} (\bibinfo{year}{2011}).

  
  \bibitem{Griesmaier_2005}A. Griesmaier, J. Werner, S. Hensler, J. Stuhler, and T. Pfau,
Phys. Rev. Lett. 94, 160401 (2005).

\bibitem{Griesmaier_2006}A. Griesmaier, J. Stuhler, T. Koch, M. Fattori, T. Pfau, and
S. Giovanazzi, Phys. Rev. Lett. 97, 250402 (2006).

\bibitem[{\citenamefont{Ni et~al.}(2010)\citenamefont{Ni, Ospelkaus, Wang,
  Qu{\'e}m{\'e}ner, Neyenhuis, De~Miranda, Bohn, Ye, and Jin}}]{ni2010dipolar}
\bibinfo{author}{\bibfnamefont{K.-K.} \bibnamefont{Ni}},
  \bibinfo{author}{\bibfnamefont{S.}~\bibnamefont{Ospelkaus}},
  \bibinfo{author}{\bibfnamefont{D.}~\bibnamefont{Wang}},
  \bibinfo{author}{\bibfnamefont{G.}~\bibnamefont{Qu{\'e}m{\'e}ner}},
  \bibinfo{author}{\bibfnamefont{B.}~\bibnamefont{Neyenhuis}},
  \bibinfo{author}{\bibfnamefont{M.}~\bibnamefont{De~Miranda}},
  \bibinfo{author}{\bibfnamefont{J.}~\bibnamefont{Bohn}},
  \bibinfo{author}{\bibfnamefont{J.}~\bibnamefont{Ye}}, \bibnamefont{}
  \bibinfo{author}{\bibfnamefont{D.}~\bibnamefont{Jin}},
  \bibinfo{journal}{Nature} \textbf{\bibinfo{volume}{464}},
  \bibinfo{pages}{1324} (\bibinfo{year}{2010}).
  
\bibitem{Cunden_2017} F. D. Cunden, P. Facchi, M. Ligab\`o, P. Vivo, J. Stat. Mech. {\bf P053303} (2017).

\bibitem{Cunden_2018} F. D. Cunden, P. Facchi, M. Ligab\`o, P. Vivo,  
J. Phys. A: Math. Theor. {\bf 51}, 35LT01 (2018). 

\bibitem{Bouchaud_2015}J-P. Bouchaud and M. Potters, {\it The Oxford Handbook of Random Matrix Theory}, The Oxford Handbook of Random Matrix Theory, 
Edited by G. Akemann, J. Baik, and P. Di Francesco, Oxford University Press, (2015) DOI: 10.1093/oxfordhb/9780198744191.013.40

\bibitem{Qiu_2017}R. C. Qiu, P. Antonik, {\it Smart Grid using Big Data Analytics: A Random Matrix Theory Approach}, John Wiley \& Sons, (2017)

\bibitem{im1} A. G. Abanov, A. Gromov, M. Kulkarni, J. Phys. A: Math. Theor. 44 ,295203 (2011)

\bibitem{im2} M. Kulkarni, A. P. Polychronakos, J. Phys. A: Math. Theor. 50 455202 (2017)

\bibitem{im3} A. K. Gon, M. Kulkarni, J. Phys. A: Math. Theor. 52, 415201 (2019) 


  
  
  \bibitem[{\citenamefont{Brown and Carrington}(2003)}]{brown2003rotational}
\bibinfo{author}{\bibfnamefont{J.~M.} \bibnamefont{Brown}} \bibnamefont{}
  \bibinfo{author}{\bibfnamefont{A.}~\bibnamefont{Carrington}},
  \emph{\bibinfo{title}{Rotational spectroscopy of diatomic molecules}}
  (\bibinfo{publisher}{Cambridge University Press}, \bibinfo{year}{2003}).
  
  
\bibitem{Tang_2018}Yijun Tang, Wil Kao, Kuan-Yu Li, Sangwon Seo, Krishnanand Mallayya, Marcos Rigol, Sarang Gopalakrishnan, and Benjamin L. Lev 
Phys. Rev. X 8, 021030, (2018).

\bibitem{Jones_2017}M P A Jones, L G Marcassa and J P Shaffer, J. Phys. B: At. Mol. Opt. Phys. 50 (2017) 060202

\bibitem{Polychronakos_1992b}Polychronakos A P,  A new integrable system with a quartic potential Phys. Lett. B {\bf 276} 341-6, (1992).

\bibitem{Garraway_2016}Barry M Garraway and H\'el\`ene Perrin, J. Phys. B: At. Mol. Opt. Phys. {\bf 49} 172001 (2016).

\bibitem{Polychronakos_1992a}Polychronakos A P,  New integrable systems from unitary matrix models Phys. Lett. B {\bf 277} 102-8, (1992).

\bibitem{Gaunt_2013}Gaunt A L, Schmidutz T F, Gotlibovych I, Smith R P and Hadzibabic Z, Bose-Einstein
condensation of atoms in a uniform potential Phys. Rev. Lett. {\bf 110} 200406, (2013).

\bibitem{box1} T. F. Schmidutz, I. Gotlibovych, A. L. Gaunt, R. P. Smith, N. Navon and Z. Hadzibabic, Phys. Rev. Lett. 112, 040403 (2014). 

\bibitem{box2} N. Navon, A. L. Gaunt, R. P. Smith and Z. Hadzibabic, Science 347, 167 (2015). 

\bibitem{box3} S. J. Garratt, C. Eigen, J. Zhang, P. Turzák, R. Lopes, R. P. Smith, Z. Hadzibabic, and N. Navon, Phys. Rev. A 99, 021601(R) (2019). 

\bibitem{Pandey_2017}Akhilesh Pandey, Avanish Kumar, and Sanjay Puri, Phys. Rev. E 96, 052211, (2017).

\bibitem{Pandey_2019a} Akhilesh Pandey, Avanish Kumar, and Sanjay Puri, arXiv:1905.10524

\bibitem{Pandey_2019b} Akhilesh Pandey, Avanish Kumar, and Sanjay Puri, arXiv:1905.10530

\bibitem{Dean_2006}D.S. Dean and S.N. Majumdar, Phys. Rev. Lett., 97, 160201 (2006).


\bibitem{Dean_2008} D. S. Dean and S. N. Majumdar, Phys. Rev. E {\bf 77}, 041108 (2008).

\bibitem{Lahiri_2000}R. Lahiri, M. Barma, and S. Ramaswamy,  Phys. Rev. E 61, 1648, (2000).

\bibitem{Allez_2012}R. Allez, J-P Bouchaud, and A. Guionnet, Phys. Rev. Lett. 109, 094102  (2012).

\bibitem{Allez_2013}R. Allez, J.-P. Bouchaud, S.N. Majumdar, P. Vivo, J. Phys. A.: Math. Theor. 46, 015001 (2013).


\bibitem{qubit} Zhang, Jiehang, Guido Pagano, Paul W. Hess, Antonis Kyprianidis, Patrick Becker, Harvey Kaplan, Alexey V. Gorshkov, Z-X. Gong, and Christopher Monroe. "Observation of a many-body dynamical phase transition with a 53-qubit quantum simulator." Nature 551, no. 7682 (2017): 601-604.

\bibitem{srep} L. L. Yan, W. Wan, L. Chen, F. Zhou, S. J. Gong, X. Tong and M. Feng. "Exploring structural phase transitions of ion crystals." 
Scientific Reports {\bf 6}, Article number, 21547 (2016)
 
\bibitem{dp} Klumov, B. A. "On the Effect of Confinement on the Structure of a Complex (Dusty) Plasma." JETP Letters 110.11 (2019): 715-721


\bibitem{Kulkarni_2012}Manas Kulkarni and Alexander G. Abanov, Phys. Rev. A 86, 033614 (2012).

\bibitem{Spohn_2014}Herbert Spohn, J Stat Phys (2014) 154, 1191-1227

\bibitem{Dean_1996}David S Dean 1996 J. Phys. A: Math. Gen. 29 L613

\bibitem{Demery_2014}Vincent D\'emery, Olivier B\'enichou and Hugo Jacquin, New Journal of Physics, 16(2014) 053032.


\bibitem{DM_2006} D. S. Dean and S. N. Majumdar, Phys. Rev. Lett. {\bf 97}, 160201 (2006).


\bibitem{Kundu_2016}A Kundu, J Cividini, EPL (Europhysics Letters) 115 (5), 54003, (2016).

\bibitem{Dhar_2018}A Dhar, A Kundu, SN Majumdar, S Sabhapandit, G Schehr
Journal of Physics A: Mathematical and Theoretical 51 (29), 295001, 2018.



\end{thebibliography}
\end{document}